\documentclass[a4paper,11pt]{article}
\usepackage{amsmath}
\numberwithin{equation}{section}
\usepackage{mathrsfs}
\usepackage{amssymb}
\usepackage{accents}
\usepackage{extarrows}
\usepackage{makecell}
\usepackage{url}
\usepackage{hyperref}
\hypersetup{colorlinks,
            citecolor=blue,
            linkcolor=blue,
            urlcolor=green,
            pdftex}

\begin{document}
\title{The Analysis to Quasi-Local Energy and Hamiltonian Constraint based on Variation}

\author{Qian. Chen\footnote{chenqian.phys.2010@gmail.com}\\ \textit{School of Physics, Shandong University, Jinan, Shandong, China}}

\date{}

\maketitle{}

\begin{abstract}
In this paper, by arising condition in variation, from equal time to non-equal time, I reconsider how geometrodynamics equations allow to be derived from variational principle in general relativity and then find the variation of extrinsic curvature dependent only locally on its induced metric and unit normal. I thus try to attribute the quasi-local energy to the integrability of submanifold. At last I discuss the dynamical degrees of freedom on Hamiltonian constraint by analyzing non-equal time variation which also represents a global transformation.
\end{abstract}

\section{Introduction}

The relationship between action and Hamiltonian have been applied to most of physics theories including constraints system. The general relativity is a famous example. In the 1962, Arnowitt, Deser and Misner explicitly discussed the dynamics of fields with generating function \cite{Arnowitt2004}, that is another theme in this paper. I mean to eliminate the momentum constraint entirely by variation, so is eliminated also in Poincar\'e-Cartan formalism. They proved the impossibility to localize the Energy for Gravitational field. I do not mean to localize it as well. In order to obtain Einstein equation from variational principle, Gibbons and Hawking took account of extrinsic curvature into gravitational action to cancel the metric derivative which straightly related to quasi-local energy in Brown and York's research by Hamilton-Jacobi analysis under orthogonal boundary in \cite{Brown1993}. Non-orthogonal boundaries were considered more carefully after Hayward put corner term into gravitational action in \cite{Hayward1993}, for instance, in \cite{Booth1999}, Booth and Mann focused, by analysing the corner term, on the leaf orthogonal to $\mathcal B$ locally rather than the $\Sigma_t$. Brown, Lau and York presented a detail mathematics to clarify the boost relationship and then transformation between different choices of foliation by it in \cite{Brown2002}.

Here is table indicating main definition in this paper.\\*
\begin{tabular}{cccccc}
\hline\hline
Manifold&Metric&\makecell{Covariant\\derivative}&\makecell{Unit\\normal}&\makecell{Intrinsic\\curvature}
&\makecell{Extrinsic\\curvature}\\
\hline
Spacetime $\mathcal M$&$g_{ab}$&$\nabla$& &$\Re_{abcd}$& \\
\\
\makecell{Hypersurfaces $\Sigma_t$\\embedded in $\mathcal M$}& \makecell{\\$h_{ab}$} & \makecell{\\$D$} & \makecell{\\$n_a$} & \makecell{\\$R_{abcd}$} & \makecell{\\$K_{ab}$}\\
\\
\makecell{Three-boundary $\mathcal B$\\satisfied $\mathcal B=\partial\mathcal M-\Sigma_{t_1}-\Sigma_{t_0}$}&\makecell{\\$\gamma_{ab}$}&\makecell{\\$\mathcal D$}&\makecell{\\$u_a$}& &\makecell{\\$\Theta_{ab}$}\\
\\
\makecell{Two-boundary $\partial\Sigma_t$\\embedded in $\Sigma_t$}&\makecell{\\$\sigma_{ab}$}&\makecell{\\$\tilde{D}$}&\makecell{\\$r_a$}& &\makecell{\\$k_{ab}$}\\
\hline\hline
\\
\end{tabular}
Where define metric tensor $h_{ab}=g_{ab}+n_a n_b$, $\gamma_{ab}=g_{ab}-u_a u_b$, $\sigma_{ab}=h_{ab}-r_a r_b$
and the extrinsic curvature tensor $K_{ab}=h_a^c h_b^d \nabla_c n_d$ of $\Sigma_t$, $\Theta_{ab}=\gamma_a^c\gamma_b^d\nabla_c u_d$ of $\mathcal B$, $k_{ab}=\sigma_a^c \sigma_b^d D_c r_d$ of $\partial\Sigma_t$ embedded in $\Sigma_t$, respectively. The spacetime ADM decomposition (see \cite{Arnowitt2004}) writes as $t^a=V^a+Nn^a$, where $t^a$ is time-flow vector field, $N$ is lapse function, $V^a$ is shift vector field, respectively.

The original motive of this paper is to examine where the baseline is that allows geometrodynamics equations derived from variation principle and then get the Hamiltonian, the most originally in all of them, to study whether the corner term is removable or not. Always appears in variation of Trace-K action, the $\displaystyle\int_{\partial\Sigma_{t_0}}^{\partial\Sigma_{t_1}}\theta\delta\sqrt{\sigma}d^2x$ (see \cite{Booth1999,Brown2002,Hayward1993}) implies that the $\theta$ is a momentum conjugated $\sqrt{\sigma}$. In fact, it is more like a condition to constrain the manifold $\mathcal B$ which associates with configurations on boundary that (see \cite{Hawking1996})
\[
\tanh\theta=-\frac{V^cr_c}{N}
\]
$\delta\theta$ is assembled by variation of configurations, moreover, no equation with respect to this quantity. I intend to take it as a momentum from manifold $\mathcal B$ itself. As far as this viewpoint are considered, there exists numerous choices for Hamiltonian, yet the manifold $\mathcal B$ selects the quasi-local energy from those Hamiltonian.

The another subject is about dynamical quality of Hamiltonian constraint, I concentrate on its relevancy with time evolution by invariance in variation formalism, in fact, this analysis through all this paper.

Section \ref{2.VP} presents how Hamiltonian arises from non-equal time variation which we are able to take it as a global transformation in general (see \cite{Arnowitt2004}), then shows its relationship with Hamilton-Jacobi analysis and Poincar\'e-Cartan integral invariance. Next I prepare a variation charge in subspace for gravitational action from the consideration about integrability where the action will be rewritten in dynamics, not geometry. Following this charge, any induced intrinsic quantities in its subspace vary as if it were not embedded in a higher dimensional manifolds, a little similar with ideal Lie algebra. I think this charge reasonable since I am not sure if the Hilbert action embedded in a higher dimensional manifolds as well. This requirement also holds the integrability about those intrinsic quantities. Yet about extrinsic curvature this rule would invalid for it described partly by its normal space, specially, exists a question about equivalence of both views in variation that $\partial\Sigma_t$ with respect to $k$ are embedded in $\Sigma_t$ or locally in $\Sigma_t\bigcup\vec n$. The subsection \ref{sub.Afbs} details this question and proves the equivalence that the variation of extrinsic curvature's contraction can still treat spanned by submanifold $\Sigma_t$ as if it were entire space. In this proof I use a condition which originates directly from non-equation variation need, otherwise the variational configurations fixed in $\partial\Sigma_t$ in equal time variation.

In section \ref{3.D} the gravitational action decomposed into ``dynamical'' formalism will be proven suitable for geometrodynamics equations, a little differentia from common ``geometric'' formalism in sense. An argument based on subsection \ref{sub.Afbs} remarks the choice of leaf unrestrictive hence its respectively extrinsic curvature in ``dynamical'' formalism (see \cite{Arnowitt2004,Ashtekar1988}). An extreme case is to choose $\mathcal B$ as ``spatial'' boundary where boundary term canceled, in fact, that is a timelike region without physical sense. Next I try to get a reason, inspires very much from \cite{Booth1999}, by importing integrability to select a Hamiltonian as quasi-local energy (since had \cite{Arnowitt2004} proved the impossibility to localize energy in general relativity.) from numerous Hamiltonian derived from unrestrictive action. The 2-form $dr_{ab}$ represents the compatibility in integrability of locally manifold with actual $\mathcal B$ then attempt to explain why action formalism unrestrictive. However, this attempt needs more replenishment.

The next section applies non-equal time variation to analyze the dynamics. Of course any variational action could be write as two parts that one of part is non-equal time variation on ''time'' boundary and another is integration whose interior is equal time variation, and the treatment usually needs $\delta q=\delta_0 q+\dot q\delta t$ (ie: $\delta_0$ denotes equal time variation, the independent variation in \cite{Arnowitt2004}), but we will hard to see what variables change under global transformations exactly then there may exists possibility to reduce variation formalism for some variables.

Section \ref{4.THC} remarks that evolution of $N\mathcal H$ multiplied by $\delta t$ in variation represents a ``true dynamical degrees of freedom'' in \cite{Wald1984}, for correlation with time and as evolution equation itself. The Dirac conjecture and re-parametrization are associated by $\pounds_{\vec t}\left(N\mathcal H\right)$ in variation formalism, which also links to Poincar\'e-Cartan integral invariance. Subsection \ref{sub.Dc} presents details to analyze Dirac conjecture from these different formalism. Subsection \ref{sub.R-p} is re-parametrization for general relativity, ``Already parameterized theory'' (see \cite{Arnowitt2004}). The Hamiltonian constraint vanishes from configurational action variation formalism after parameter transformed in case of particle action, yet still appears in case of gravitational action. In case of the former, the constraint vanishes by Legendre transformation, regardless of the system obeys Lagrangian equations or not. Yet, in case of the latter, only the system obeys Lagrangian equations the constraint equal to zero. Indeed, configurational action always possesses the variation which have removed the Hamiltonian constraint generated by parametrization hence frankness in this variation formalism. The variation of canonical action reveals this characteristic as well though a little complex. I only consider re-parametrization which is most analogous to particle action in physical sense, that is to view lapse function as the transformation coefficient, the source of re-parametrization ideal as well. The analysis recovers the former research (see \cite{Ashtekar1988}) conclusions that both of Dirac conjecture and de-parametrization are invalid in gravitational field. Thus I guess the $\mathcal H$ is both of constraint and true Hamiltonian. In this section I refer nothing about first-class or second-class constraint.

\section{Variation Principle} \label{2.VP}
\subsection{Varying Velocity}
The functional variation can view as parameter group of diffeomorphism thereby we are able to compute by Lie derivative. Consider a one-parameter group of diffeomorphism described with parameter $\lambda$. Any functional variation induced by it define incidentally Lie derivative with respect to vector $\vec \lambda$. From the definition of pull back we have $d(\pounds_{\vec \lambda} f)=\pounds_{\vec \lambda} df$.

Functional variation of velocity can express as commutator of velocity parameter and variation parameter, that is, \begin{equation}
\pounds_{\vec \lambda} \pounds_{\vec t} f-\pounds_{\vec t} \pounds_{\vec \lambda} f=[\vec \lambda , \vec t\,](f)
\end{equation}
After some algebra, we obtain
\begin{equation}
\pounds_{\vec \lambda}\left(\frac{\partial}{\partial t}\right)=\lim_{\lambda \to 0} \frac{\frac{\partial}{\partial t(\lambda)}-\frac{\partial}{\partial t(0)}}{\lambda}=-\frac{d\pounds_{\vec \lambda}t}{dt} \frac{\partial}{\partial t}
\end{equation}

According to above argument, we may replace the variation with the Lie derivative which usually symbolizes a global transformation, and a more emphatic aspect, any variational terms in satisfy Leibnitz rule including derivative, so just evaluate from one part to another even in those difficult case.

\subsection{Non-equal time variation} \label{s:ntc}
It is common to cancel $\delta t$ during variation process for convenience. Let us call it non-equal time variation when take $\delta t\neq0$ into consideration\footnote{in \cite{Arnowitt2004}, we can see the $\delta q=\delta_0 q+\dot q\delta t$. The manner which $\delta I=\delta I_0+L\delta t$ and $\delta I_0$ is equal to zero for the classical solutions may neglect some terms which could be absorbed in $\delta I_0$.}, which we have known as non-commutative between variation and differentiation of the generalized velocity. Variational action with respect to Lagrangian $dI=Ldt$ writes as
\begin{equation}
\delta (dI)=(\delta L)dt+Ld(\delta t)
\end{equation}
Split $\delta (dI)$ to familiar form, via Legendre transformation $p=\frac{\partial L}{\partial \dot q}$ and $H=p \dot q-L$, then obtain
\begin{equation}
\delta (dI)=\left(\frac{\partial L}{\partial q}-\frac{d}{dt} \frac{\partial L}{\partial \dot q}\right)\delta q dt+d(p \delta q)-Hd(\delta t)+\frac{\partial L}{\partial t} \delta t dt
\end{equation}
Integrate by the $\delta (dI)=d(\delta I)$
\begin{equation} \label{eq:ntc}
\delta I=p \delta q -H \delta t+\int\left(\frac{\partial L}{\partial q}-\frac{d}{dt} \frac{\partial L}{\partial \dot q}\right)\delta q dt+\left(\frac{dH}{dt}+\frac{\partial L}{\partial t}\right) \delta t dt
\end{equation}

Not difficult to find that the first term on the right side recover the conclusion of Hamilton-Jacobi analysis (see \cite{Arnold1989,Brown1993}), and we are able to view it as a generating function (see \cite{Arnowitt2004})
\[
\delta I(\text{classical solutions})=\mathcal G=p\delta q-H\delta t
\]
More emphatic, $\delta q$, $\delta t$ are arbitrary, so is independent to each another. The Hamiltonian arises from the coefficient of $\delta t$. Conversely, the coefficient of $\delta t$ should be viewed as the Hamiltonian.

Primary constraints always come out from Legendre transformation thus absent in variational action formalism if we evaluate the components of the functional variation one by one. Yet secondary constraints exist in variation formalism since they correspond the Lagrangian equations with respect to primary constraints.

We are able to look on (\ref{eq:ntc}) as a relativistic conclusion since the Hamiltonian $H$ can be viewed as a momentum conjugated $t$ which can be interpreted as configuration, by Legendre transformation. If we substitute $\tau$ for $t$ as time parameter, it will recover the (\ref{eq:ntc}). The proof that define the $dt=Nd\tau$, and put it into the action then it rewrites as $\displaystyle I=\int\!NLd\tau$. After straightforward calculation, it is easy to obtain (\ref{eq:ntc}) once again, and very notable, $\tau$ vanished completely by Legendre transformation. The process of replacing $t$ with $\tau$ corresponds to the re-parametrization in Hamiltonian theory that we call the $t$ as a coordinate and $\tau$ as a new dynamical parameter yet this treatment generates a Hamiltonian constraint for Hamiltonian $H_{\tau}$ from Legendre transformation, an identity $H_{\tau}=0$ at all over configurational space therefore it is removable from variation formalism.

The formula (\ref{eq:ntc}) also leads to Poincar\'e-Cartan integral invariant (see \cite{Arnold1989}). Actions of two neighbor trajectories which obey canonical equations at the phase space satisfy
\begin{displaymath}
\delta I\Big|_{i}^{i+1}=p(t_1(i))\delta q(t_1)\Big|_{i}^{i+1}-H(t_1(i))\delta t_1\Big|_{i}^{i+1}-p(t_0(i))\delta q(t_0)\Big|_{i}^{i+1}+H(t_0(i))\delta t_0\Big|_{i}^{i+1}
\end{displaymath}
Here $t_0$ and $t_1$ are the initial and the final points of any trajectories and the ${i}$ label trajectories. Note the $\delta q(t_0)\Big|_{i}^{i+1}\equiv q(t_0(i+1))-q(t_0(i))$, then integrate $\delta I$
\begin{displaymath}
\oint_{C_0}^{C_1} p\delta q-H\delta t=\delta I\Big|_{n}^{1}+\lim_{n \to \infty}\sum_{i=1}^{n-1} \delta I\Big|_{i}^{i+1}=0
\end{displaymath}
The integration goes over any closed circuit in phase space which each trajectory crosses this circuit only once. The equation shows that
\begin{equation}
\oint_{C}p\delta q-H\delta t=constant
\end{equation}
Where $C$ is any one of closed circuit on trajectory tube $\cup_tC_0$ to which any one of closed circuit $C_0$ on phase space give rise by canonical equations, namely, by time-flow field $\vec t$.

\subsection{Tensor density} \label{sub.Td}
In order to analyze action of field theory, the Lie derivative of tensor density $\accentset{\smallfrown}{\mathscr L}$ may well be a conventional manner that the formulation expresses as
\begin{displaymath} \label{density}
\pounds_{\vec \lambda}\left( \accentset{\smallfrown}{\mathscr L}{\scriptscriptstyle\circ} \bar{\boldsymbol{e}}\boldsymbol{e} \right)=\boldsymbol{e}\pounds_{\vec \lambda}\mathscr L+m\boldsymbol{e}\mathscr L\partial_a\lambda^a
\end{displaymath}
Where the $\boldsymbol{e}$, $\bar{\boldsymbol{e}}$ denote the volume element of coordinate basis and there is relationship $\bar{\boldsymbol{e}}\boldsymbol{e}=4!$ . Specially, A Lagrangian possesses tensor densities wight of $m=1$, (\ref{density}) rewrites by $\partial$ (see \cite{Arnowitt2004})
\[
\delta\mathscr L=\delta_0 \mathscr L+\frac{\partial\left(\mathscr L \delta x^{\mu}\right)}{\partial x^{\mu}}
\]
A scalar multiplying $\sqrt{-g}$ can be thought of as its dedensitization. For example, variation of Hilbert action is
\begin{displaymath}
\pounds_{\vec \lambda}(\Re\sqrt{-g})=\sqrt{-g}\lambda^a\partial_a\Re+\Re\lambda^a\partial_a\sqrt{-g}
\end{displaymath}
Put $\boldsymbol{e}\mathscr L\partial_a\lambda^a$ into (see \cite{Brown2002})
\begin{displaymath}
\pounds_{\vec \lambda}(\Re\sqrt{-g}\boldsymbol{e})=\sqrt{-g}\boldsymbol{e}(\pounds_{\vec \lambda}\Re+\frac12\Re g^{ab}\pounds_{\vec \lambda}g_{ab})
\end{displaymath}
The consistency in form with variational operator $\delta$ let us be able to express the global transformation as variational action.

\subsection{Decompose variational vector} \label{DI}
On the basis of summing definition, an integration must be independent on the choice we divide interval or region. We can suppose a variation is purely inner that all of variational vector fields with respect to Lie derivative are spanned by integral region itself which contributes nothing to total integration hence the variation equal to zero. \cite{Brown2002} discuss a example of this kind of variation and call it diffeomorphism invariance, very appropriate.

Noting that the vector $\vec \lambda$ not only lie along manifolds but configuration space even fibre bundles\footnote{Well, I mean all the derivatives with respect to configurations, for instance, velocity.}, we have $\vec \lambda=\vec \zeta+\vec \xi$, where the $\vec \zeta$ denotes the vector along configuration and fibre bundles space yet the $\vec \xi$ along manifold. The variation will become boundary term if $\vec\zeta=0$ and $\vec\xi$ has normal component on boundary. It is a way to examine whether the expression of variation is right or not. For a instance, the variation of particle action is (\ref{eq:ntc}), now let it be purely inner variation, we have $\delta t=\pounds_{\!\vec \xi}\,t$, then
\[
\delta I_{\mathrm{i}}=\int\delta t\left[ \dot q\left(\frac{\partial L}{\partial q}-\dot p\right)-\left(\frac{\partial L}{\partial t}+\frac{dH}{dt}\right)\right]dt=0
\]
Here Legendre transformation $H=p\dot q-L$ applied. Similarly, the inner variation of Hilbert action without boundary term is
\begin{displaymath}
\delta I_{\mathrm{i}}=-2\int_\mathcal M G_{ab}\nabla^a \xi^b d^4x=2\int_\mathcal M \xi^b\nabla^aG_{ab}d^4x=0
\end{displaymath}
Where the boundary term has been neglected at the second equals sign.

The decomposition $\vec\lambda=\vec\zeta+\vec\xi$ divides any action as equal time variation and non-equal time variation and it derives a condition for variation on the boundary $\partial\Sigma_t$ spanned by $\Sigma_t$ if this action is attributed to those integration which are integrated by integration over foliation with arising by time flowing
\begin{equation*} \label{c1}
\delta q \Big|_{\partial \Sigma_t}=\pounds_{\vec \xi}\,q,~~~\forall t\in [t_0, t_1] \tag{i}
\end{equation*}
Here vector $\vec\xi=\xi\vec t$ then $\xi=\delta t$. Validate the condition only when its targets can be attributed to configurational variables variation without fibre bundles, therefore it is better that we get the variation before using the condition than using it directly. Condition (\ref{c1}) means to continue non-equal time variation on $\partial\Sigma_t$ by appointing $\vec\lambda$'s decomposition since configurations fixed if considers equal time variation merely.

\subsection{Integrability for variation} \label{integrable}
The geodesic problem is one of one-dimension variational problem. The length of a curve on manifold $\mathcal M$ with metric $g_{ab}$ expresses as
\begin{displaymath}
l=\int\!\sqrt{g\left( \vec t, \vec t \:\right)}dt
\end{displaymath}
Let $\phi\colon\lambda \times \mathcal M \to \mathcal M$ be a one-parameter group of diffeomorphisms as the variational group. Now define the variation of any function is
\[
\lim_{\lambda \to 0}\frac{\phi_{\lambda}^*f-f}{\lambda}=\lim_{\lambda \to 0}\frac{f(\lambda)-f(0)}{\lambda}
\]
After straightforward manner to vary the length, obtain
\begin{displaymath}
\delta l=\int\!\frac{t^at^b\delta g_{ab}}{2\sqrt{g\left( \vec t, \vec t \:\right)}}dt+\int\Bigl[ \frac{t^b g_{ab}\delta t^a}{\sqrt{g\left( \vec t, \vec t \:\right)}}dt+\sqrt{g\left( \vec t, \vec t \:\right)}d(\delta t) \Bigr]
\end{displaymath}
The first integral contains the geodesic equation. Note that both $\phi_{\lambda}(L)$ and $L$ are integrable curves, thus, we are able to prove the second integral vanishes since the pull back requires the pull back of a tangent vector of $\phi_{\lambda} (L)$ to be a vector tangent to $L$, in other words, the integrability of $\phi_{\lambda}^* \vec t$ is comparable with $L$. The proof is direct for $\delta t^a=\alpha t^a$ and $t^a\nabla_at=1$.

Let us extend the rule of variation associated with integrable. For an integral on manifold $M$, the rule shall become
\begin{equation} \label{vc}
\delta\int_{\mathcal M}\omega _n\Bigl( \mathscr F_{\mathcal M}, \nabla \mathscr F_{\mathcal M} \Bigr)=\int_{\mathcal M}\omega_n'\Bigl( \mathscr F_{\mathcal M}, \nabla \mathscr F_{\mathcal M}, \delta\mathscr F_{\mathcal M}, \nabla \overline{\delta\mathscr F_{\mathcal M}} \Bigr)
\end{equation}
Where the $\mathcal M$ represents any given n-dimensional manifold including submanifold embedded in a bigger space and the $\omega_n$ indicates n-form. The $\nabla \mathscr F_{\mathcal M}$ denote the tensor generated by any given tensor $\mathscr F_{\mathcal M}$ intrinsic for $\mathcal M$, with n-dimensional derivative operator $\nabla$ with respect to $\mathcal M$. Furthermore, we define a connection tensor ${}^nC$ generated by varying $\nabla$
\[
\delta\,\nabla={}^nC\left( \mathscr F_{\mathcal M},\overline{\delta\mathscr F_{\mathcal M}},\nabla \overline{\delta\mathscr F_{\mathcal M}} \right)
\]
Then it is absorbed in $\omega_n'$. For instance, foliation $\Sigma_t$ is submanifolds embedded in $\mathbb{R}^4$, and each leaf curvature tensor is intrinsic quality. The connection generated by variation is defined as
\begin{equation} \label{rvc}
D_aw_b-\overset{\lambda}{D}_aw_b={}^3C^c_{\phantom{c}ab}w_c, \:\:\:\:\:w_b\in W_{\Sigma_t}^*
\end{equation}
Furthermore, we have
\begin{equation} \label{vr}
{}^3C^c_{\phantom{c}ab}=\frac12h^{cd}\Bigl( 2D_{(a}\overline{\delta h_{b)d}}-D_d\overline{\delta h_{ab}} \Bigr)
\end{equation}

It is rational that variation of intrinsic quantities in subspace holds intrinsic, a little similarity with ideal Lie algebra. Conversely, rule (\ref{vc}) loses its validity when used in extrinsic curvature since it must apply higher dimensional derivative operator to the normal vector of the submanifold. However, we can still expect that the variation in submanifold within an ``large'' submanifold equals to in submanifold, this point is discussed in \ref{sub.Afbs}.

In the case of Trace-K action (see \cite{Brown2002}), as far as the condition (\ref{c1}) is considered, the unit normal $u_a$ of $\mathcal B$ vectored by variation operator $\delta$ have relationship between any $v^a$ tangent $\mathcal B$ that
\[
v^a\delta u_a=-u_a\delta v^a=-u_a[\vec \lambda, \vec v\,]^a=0
\]
Here using the Frobenius's theorem (see \cite{Wald1984}) and noting $\delta u_a$ orthogonal to $\mathcal B$. We thus have $\delta u_a=\frac12u_au^bu^c\delta g_{ab}$ if the $u^au_a=1$, one of two identities in the lemma about varying hypersurface-orthogonal dual vector in \cite{Brown2002}. On the other hand, for those Lagrangian based on foliation, it is natural to find that the foliation are held if view each leaf still as a entirety under variation, which we are able to imagine as $\delta t$ between any two leaves are equal on every points for both of one, namely, $D_a\delta t=0$, or
\begin{equation*} \label{c2}
h_a^c\delta n_c=0 \tag{ii}
\end{equation*}
In summary the formula $\delta u_a=\frac12u_au^bu^c\delta g_{bc}$ (see \cite{Brown2002}) is established if the maps they generate hold foliation integrable.

\subsection{Independence on boundary embedded} \label{sub.Afbs}
Suppose the boundary possessing $k$, the extrinsic curvature of $\partial\Sigma_t$ embedded in $\Sigma_t$, we may structure an augmented space $\partial\Sigma_t\bigcup\vec n$ locally to make it higher dimensions without changing the normal vector $r^a$ and extrinsic curvature $k$. The argument exhibit also from the definition of $k$ and the operator $D$, that is
\[
k=\sigma^{ab}\sigma_a^c\sigma_b^dD_cr_d=\sigma^{ab}\sigma_a^c\sigma_b^d\nabla_cr_d
\]
Thus, there is no contribution to extrinsic curvature in value though operators divergence in dimensionality. Now we must prove the consistency between $\partial\Sigma_t\bigcup\vec n$ and $\partial\Sigma_t$ in variation. Obviously for the result of $\partial\Sigma_t$, thus consider $\partial\Sigma_t\bigcup\vec n$, that is
\begin{equation} \label{vk} \begin{split}
2\delta k=&k_{ab}\delta\sigma^{ab}+\tilde{D}_c\left(\sigma_a^c\delta r^a\right)+\sigma^{ab}\nabla_a\left(\sigma_b^c\delta r_c\right)+\left(\sigma_c^br^a-\sigma^{ab}r_c\right){}^3C^c_{\phantom{c}ab}\\
&-2K^{ab}r_a\delta n_b+2Kr^c\delta n_c+2h_a^b\delta n_bn^c\nabla_c r^a-2\delta r_ar^cD_cr^a\\
\end{split} \end{equation}
Here we employ the $\overline{C^c_{\phantom{c}ab}}={}^3C^c_{\phantom{c}ab}-h^{cd}n^eK_{ab}\delta g_{de}$. Using the condition (\ref{c1}) hence $\delta r_c=\pounds_{\xi \vec t}\,r_c$, and importing (\ref{c2}), we obtain $\sigma_a^c\delta r_c=0$ where imitate the argument in \ref{integrable}, that the pull back of a vector field tangent $\partial\Sigma_t$, which is laying certain curve, will be still tangent along its mapping curve. Now rewrite (\ref{vk}) as
\begin{equation} \label{vk1}
2\delta k=k_{ab}\delta\sigma^{ab}+\tilde{D}_c(\sigma_a^c\delta r^a)+\left(\sigma_c^br^a-\sigma^{ab}r_c\right){}^3C^c_{\phantom{c}ab}
\end{equation}
This proof demonstrates $\delta k$ disposed locally in action based on foliation. We will see its significance in \ref{sub.Aiaf}.

\section{Dynamics} \label{3.D}
\subsection{``Dynamical'' action}
Apply the ADM \cite{Arnowitt2004} decomposition $t^a=V^a+Nn^a$ to the scalar curvature
\begin{displaymath} \begin{split}
\Re &=R+K^2-K_{ab}K^{ab}-2R_{ab}n^a n^b \\
&=R+K_{ab}K^{ab}-K^2+2[\nabla_a(n^a \nabla_c n^c)-\nabla_c(n^a \nabla_a n^c)]
\end{split} \end{displaymath}

The $\nabla$ terms may rewrite as extrinsic curvature by Gauss law, however, to say strictly, a smooth boundary spanned by $\mathcal M$ is needed for avoiding miscellaneous term. To slice the action from geometrical to dynamical, may avoid these miscellaneous terms. Indeed, we will refer the rule in \ref{integrable} for varying this foliaceous action.

Start with this action below without corner term for the sake of simplicity, that is (see \cite{Arnowitt2004,Ashtekar1988})
\begin{equation} \label{a}
I\left(q, \Sigma_t, t\right)=\int\!dt\!\int_{\Sigma_t}N(R+K_{ab}K^{ab}-K^2)\sqrt{h}d^3x+2\int\!dt\!\int_{\partial \Sigma_t}Nk\sqrt{\sigma}d^2x
\end{equation}
Where the $q$ indicates configurations $\left(h_{ab}, V^a, N\right)$ and $\Sigma_t$ represents the foliations to which we slice $\mathcal M$. The action possesses extrinsic curvature term with respect to $\partial\Sigma_t$, where its definition is
\begin{equation}
k_{ab}\equiv\sigma_a^c\sigma_b^dD_cr_d
\end{equation}

The action (\ref{a}) means a time-flow integral about Lagrangian, an integration over foliation $\Sigma_t$ where each leaf is a integrable subspace of $\mathcal M$. The variation for (\ref{a}) about each leaf is also integrable. It is not necessary to demand $\delta h_{ab}$ purely spatial, yet the projection of $\delta h_{ab}$ defined as
\[
\overline{\delta h_{ab}}=h_a^ch_b^d\delta h_{cd}=h_a^ch_b^d\delta g_{cd}
\]
We stick to demand $\delta g_a^b=0$.

Employe (\ref{vr}) and (\ref{vk1}) to vary the action (\ref{a}), then we have
\begin{equation} \begin{split} \label{vi}
\delta I=&\int dt\int_{\Sigma_t}\left[\left(\Lambda^{ab}-\dot{P}^{ab}\right)\overline{\delta h_{ab}}-\mathcal H^aNn^b\delta h_{ab}-\mathcal H\delta N+\delta t\pounds_{\vec t}\left(N\mathcal H\right)\right]d^3x\\
&+\int_{\Sigma_{t_0}}^{\Sigma_{t_1}}\left(P^{ab}\delta h_{ab}-N\mathcal H\delta t\right)d^3x+2\int_{\partial\Sigma_{t_0}}^{\partial\Sigma_{t_1}}Nk\sqrt{\sigma}\delta td^2x\\
&+\int dt\int_{\partial \Sigma_t}\sqrt{\sigma}\bigg[-\frac{2}{\sqrt{\sigma}}\delta t\pounds_{\vec t}\left(Nk\sqrt{\sigma}\right)+2k\delta N\\
&+\left(\sigma^{ab}r^cD_cN-\frac{N\pi^{ab}}{\sqrt{\sigma}}\right)\delta\sigma_{ab}
-\left(2\frac{P^{cb}}{\sqrt{h}}Nn^a+\frac{P^{ab}}{\sqrt{h}}V^c\right)r_c\delta h_{ab}\bigg]d^2x
\end{split} \end{equation}
Where applied (\ref{c2}) for $\mathcal H$, which just let us rewrite the $Nn^c\nabla_c\delta t$ as $t^c\nabla_c \delta t$, and have defined (see \cite{Arnowitt2004}, \cite{Brown1993} respectively)
\begin{align*}
\Lambda^{ab}&=\sqrt{h}\left(D^aD^bN-h^{ab}D^cD_cN\right)-N\sqrt{h}\left(R^{ab}-\frac12 Rh^{ab}\right)+D_c\left(P^{ab}V^c\right)\\
&\quad-\frac{2N}{\sqrt{h}}\left[P^{c(a}P_c^{\,b)}-\frac12 PP^{ab}\right]-2P^{c(a}D_cV^{b)}+\frac{N}{2\sqrt{h}}h^{ab}\left(P_{cd}P^{cd}-\frac12 P^2\right)\\
\pi^{ab}&=\sqrt{\sigma}\left(k^{ab}-k\sigma^{ab}\right)
\end{align*}
and the secondary constraints
\begin{align*}
\mathcal H\,&=\frac{1}{\sqrt{h}}\left(P^{ab}P_{ab}-\frac12 P^2\right)-R\sqrt{h}\\
\mathcal H^a&=-2D_bP^{ab}
\end{align*}
The Hamiltonian constraint express the arbitrariness for foliation selection that both of the action and its variation are independent of lapse function $N$ in value.

Apply (\ref{c2}) to $\mathcal H^a$, then the integral overlooked boundary term in (\ref{vi}) rewrites as
\begin{equation} \label{VG}
\int_{\mathcal M}G_{ab}\delta g^{ab}\sqrt{-g}dtd^3x
\end{equation}

Rewrite (\ref{vi}) by the condition (\ref{c1}), we have
\begin{equation} \begin{split} \label{vit}
\delta I=&\int dt\int_{\Sigma_t}\left[\left(\Lambda^{ab}-\dot{P}^{ab}\right)\overline{\delta h_{ab}}-\mathcal H^aNn^b\delta h_{ab}-\mathcal H\delta N+\delta t\pounds_{\vec t}\left(N\mathcal H\right)\right]d^3x\\
&+\int_{\Sigma_{t_0}}^{\Sigma_{t_1}}\left(P^{ab}\delta h_{ab}-N\mathcal H\delta t\right)d^3x+2\int_{\partial\Sigma_{t_0}}^{\partial\Sigma_{t_1}}\left(Nk-\frac{r_aP^{ab}V_b}{\sqrt{h}}\right)\sqrt{\sigma}\delta td^2x\\
&+\int dt\int_{\partial \Sigma_t}\delta t\sqrt{\sigma}\bigg[\frac{2}{\sqrt{\sigma}}\pounds_{\vec t}\left(\frac{r_aP^{ab}V_b}{\sqrt{h}}\sqrt{\sigma}-Nk\sqrt{\sigma}\right)+2k\dot{N}\\
&+\left(\sigma^{ab}r^cD_cN-\frac{N\pi^{ab}}{\sqrt{\sigma}}\right)\dot{\sigma}_{ab}
-2\frac{r_bP_a^b\dot{V}^a}{\sqrt{h}}-\frac{V^cr_cP^{ab}\dot{h}_{ab}}{\sqrt{h}}\bigg]d^2x
\end{split} \end{equation}

With the Legendre transformation $H=p\dot q-L$, the Hamiltonian of action (\ref{a}) is
\begin{equation} \label{h}
H=\int_{\Sigma_t}\left(N\mathcal H+V^a\mathcal H_a\right)d^3x+2\int_{\partial \Sigma_t}\left(\frac{r_aV_bP^{ab}}{\sqrt{h}}-Nk\right)\sqrt{\sigma}d^2x
\end{equation}
With conclusion of \ref{s:ntc}, the term multiplied by $\delta t$ in (\ref{vi}) is a evolutional equation for Hamiltonian. To differentiate (\ref{h}) with time directly we will have
\begin{equation} \begin{split} \label{he}
\dot H=&\int_{\Sigma_t}\left[\left(\dot{P}^{ab}-\Lambda^{ab}\right)\dot{h}_{ab}+\dot{N}\mathcal H+\dot{V}^a\mathcal H_a\right]d^3x+\int_{\partial \Sigma_t}\bigg[\frac{P^{ab}}{\sqrt{h}}V^cr_c\dot{h}_{ab}\\
&+2\frac{r_bP_a^b}{\sqrt{h}}\dot{V}^a
-2k\dot{N}+\left(\frac{N\pi^{ab}}{\sqrt{\sigma}}-\sigma^{ab}r^cD_cN\right)\bigg]\dot{\sigma}_{ab}\sqrt{\sigma}d^2x
\end{split} \end{equation}

Let substitute $\delta$ with $\vec \xi=\xi \vec t$ , the (\ref{vi}) becomes
\begin{equation} \label{ie}
\pounds_{\vec \xi}\,I=\int_{\Sigma_{t_0}}^{\Sigma_{t_1}}P^{ab}\dot{h}_{ab}d^3x-H\delta t\Big|_{t_0}^{t_1}=L\delta t\Big|_{t_0}^{t_1}
\end{equation}
(\ref{ie}) represents diffeomorphism invariance of variational action (\ref{a}) form by time flowing.

\subsection{Unrestrictive in action formalism} \label{sub.Aiaf}
Note that there are so much vector projected on $dt$ equal to $1$ which is 1-form with respect to integral, for instance $N\vec n$. The extrinsic curvature $k$ integral along time can be viewed as a process to knit $\partial\Sigma_t\bigcup\vec n$ one by one. There is no necessity to demand the three-manifold knitted by $\partial\Sigma_t\bigcup\vec n$ integrable since we are able to dispose $\delta k$ locally in dynamical action formalism as argument in \ref{sub.Afbs}. Indeed, any action with the form of (\ref{a}) that differs from $\Sigma_t$ selection at $t\in(t_0, t_1)$ can be used.

Now that variation formalism allows those action comparable with non-integrable to describe the geometrodynamics of $\mathcal M$, there are various alternate boundaries for extrinsic curvature $k$, and those choices correspond different meanings. Let us suppose an extreme situation so locally for the each leaf which makes sense on geometry but not on physics that let boundary spanned by $\Sigma_t$ approach $\mathcal B$ infinitely so that it might be thought as one of subset of the latter. Thereby, in this situation the connection term from varying $R$ where have labelled $``*"$ for distinguishing from $h_{ab}$ and $D$ writes as
\[
\left(\sigma^{ab}r_c^*-\sigma_c^br^{*a}\right){}^3C^{*c}_{\phantom{c}ab}
\]
vanishes under equal time variational condition since $\delta h_{ab}^*=0$ satisfied on each point $\mathcal B$ which leads to $h_d^{*c} D_c^*\delta h_{ab}^*=0$. The consequence of this situation seems to be an equivalence to Palatini action in \cite{Arnowitt2004,Wald1984} that it is redundant to put extrinsic curvature into action.

\subsection{Hamiltonian and quasi-local energy}
It is necessary to find other characteristic to restrict the relationship between Hamiltonian and quasi-local energy since the action formalism restricts the former insufficiently. In geometrodynamics, various vector at certain point are able to lead one leaf to another that makes the parameter $t$ altered the same. In \ref{sub.Aiaf} we have discussed the ``locally'' boundary to explain the parameter alteration led by various vector fields yet the difference of them demonstrates on augmented space $\partial\Sigma_t\bigcup\vec n$ they structure. Note that manifold $\partial\Sigma_t\bigcup\vec n$ exist locally whereas $\mathcal B$ integrable. The concept of energy always wants a quantity integrable along time. Rationally, to extend the function integrable to manifolds on which Hamiltonian are based.

As the \cite{Booth1999} had issued, it does not seem to physically make sense for the observers to measure the energy and momentum surface densities with respect to the foliation $\Sigma_t$ that is not perpendicular to $t^a$ since this foliation has not ordinary definition for space, a definition that space is perpendicular to world line vector, whereas a foliation which is perpendicular to $t^a$ has it. However, the action its boundary term $k$ always be proper toward describing the gravitational system because it would derive the geometrodynamics equations and evolution equation for boundary quantities, therefore, I am willing to attribute the energy conditions to Hamiltonian since in Hamiltonian formalism the most emphatic step is to divide the space and time. Moreover, the analysis to integrability might recover a cross term canceled by integral projection.

Consider any two vectors at $\partial\Sigma_t\bigcup\vec n$, their Lie bracket is $[\vec v^1+c_1\vec n, \vec v^2+c_2\vec n]$, here the $\vec v^1$ and $\vec v^2$ are tangent to $\partial\Sigma_t$, then, do contraction by $r_a$, the result is $r_a[\vec n, c_1v^1-c_2v^2]^a$. Noting the vector $\vec v=c_1\vec v^1-c_2\vec v^2$ tangent the $\partial\Sigma_t$ as well, we have $r_a[\vec n, \vec v]^a$, furthermore, rewrite it as $v^an^bdr_{ab}$. An integrable submanifold obeys the Frobenius's theorem, thereby the contraction of Lie bracket and $r_a$ vanish whatever any $\vec v^1$ and $\vec v^2$
\begin{equation} \label{dr}
\sigma_a^cn^bdr_{cb}=0
\end{equation}
The 2-form $dr_{ab}$ labels the integrability of a submanifold, that is another formalism about Frobenius's theorem.

Let us consider the relationship between two different foliation came from slicing the spacetime manifold $\mathcal M$ differently, here label those $\Sigma_t$ and $\Sigma'_t$, and note that the relationship about two slicing way, that is
\begin{align*}
r'^a&=\alpha r^a-\beta n^a\\
n'^a&=\alpha n^a-\beta r^a
\end{align*}
Note the $\sigma_{ab}$ invariant under the selection of foliation $\Sigma_t$ changed, and here we have defined $\alpha^2-\beta^2=1$ and $\beta=r'^an_a$. After some straightforward calculation, for any two foliation selections, we have
\begin{equation} \label{KK}
\sigma_a^cK'_{cb}r'^b=\sigma_a^cK_{cb}r^b-\sigma_a^c\nabla_c\theta
\end{equation}
Here $\sinh\theta=\beta$. Now we replace the $r'^a$ and $n'^a$ with $u^a$ and $\bar n^a$, where $\bar n^a$ is unit vector of ${}^{\mathcal B}\Sigma_t$, a choice of foliation that each normal vector on boundary spanned by it are tangent along $\mathcal B$. Only for $u^b$ we have $\sigma_a^c u^b\nabla_c\bar n_b=-\sigma_a^c\bar n^b\Theta_{bc}$. Apply (\ref{KK}) and $\sigma_a^c \bar n^b du_{cb}=0$, the relationship between the $\mathcal B$ and $\partial\Sigma_t\bigcup \vec n$ is
\begin{equation}
\sigma_a^c n^b dr_{cb}=\sigma_a^c \bar n^b\Theta_{bc}-\sigma_a^c n^b \hat k_{bc}-\sigma_a^c\nabla_c \theta
\end{equation}
The $\sigma_a^c n^b \hat k_{bc}=\sigma_a^c n^b \nabla_b r_c$ denotes the extrinsic curvature of ''locally'' hypersurfaces $\partial\Sigma_t\bigcup \vec n$. It is hard to assure the index in $\hat k_{bc}$ symmetrical except the hypersurfaces integrable. The consequence of swapping index in (\ref{KK}) equivalent to
\[
0=\sigma_a^c\bar n^b\Theta_{cb}-\sigma_a^c n^b\hat k_{cb}-\sigma_a^c\nabla_c \theta
\]
Now the $\sigma_a^c n^b dr_{cb}$ relates the difference of cross term. By straightforward calculate we have (see \cite{Booth1999})
\begin{equation}
\frac{\bar P^{ab}\bar V_a u_b}{\sqrt{\bar h}}-\bar N\bar k=\frac{P^{ab}V_a r_b}{\sqrt{h}}-Nk-V^a\sigma_a^c\nabla_c \theta
\end{equation}
We are able to see that $du_{ab}$ decides if the indexes are symmetrical in the momentum with respect to $\Theta_{ab}$ in $\cite{Brown2002}$.

The $\sigma_a^c n^b dr_{cb}$ represents the projection of $dr_{ab}$ on $\partial\Sigma_t\bigcup\vec n$, or a cross component of this 2-form with respect to $\partial\Sigma_t\bigcup\vec n$, likewise the $\sigma_a^c t^b dr_{cb}$ is the projection of $dr_{ab}$ on $\mathcal B$, and then rewrite it as
\begin{equation}
\sigma_a^c t^b dr_{cb}=\sigma_a^c \nabla_c (t^b r_b)-\sigma_a^c\pounds_{\vec t}\,r_c
\end{equation}
Where applied the $r_c=\alpha u_c+\beta\bar n_c$ and $\sigma_a^c\pounds_{\vec t}\,u_c=0$, also considered the $\pounds_{\vec t}\,\bar n_c$ canceled since $\bar n_c$ represents normal covector of foliation ${}^{\mathcal B}\Sigma_t$, thus obtain an exact 1-form $\sigma_a^c t^b dr_{cb}=\tilde D_a (t^b r_b)$ on submanifold $\partial\Sigma_t$. Now any 2-form wedge by $\sigma_a^c t^b dr_{cb}$ with an exact 1-form $d\omega$ also satisfies
\begin{equation} \label{d^d=d^}
\int_{\partial\Sigma_t} d\omega_e \wedge \sigma_a^c t^b dr_{cb}=\int_{\partial\Sigma_t} d\left(\omega_e \wedge \sigma_a^c t^b dr_{cb}\right)=0
\end{equation}

Another way to get (\ref{d^d=d^}) is by applying the manner in \cite{Brown2002} and then find $\sigma_a^c n^b dr_{cb}$ contains the angle term. Now we write
\begin{subequations}
\begin{alignat*}{2}
u_c&=\bar M\nabla_c s, &\quad r_c&=M D_c s\\
\bar n_c&=-\bar N\nabla_c t, & n_c&=-N\nabla_c t
\end{alignat*}
\end{subequations}
Where $``s"$ denotes the parameter of hypersurfaces $\mathcal B$. A relationship between $(M, \bar M, N, \bar N)$ is
\[
\alpha=\frac{\bar M}{M}=\frac{N}{\bar N}
\]
Split the $\sigma_a^cn^bdr_{cb}$ as
\begin{equation} \begin{split}
\sigma_a^c n^bdr_{cb}&=\sigma_a^c n^b\left(\nabla_c r_b-\nabla_b r_c\right)\\
&=\frac{\beta}{\alpha}\sigma_a^c\nabla_c\ln\frac{\bar M}{N}-\sigma_a^c\nabla_c\theta
\end{split} \end{equation}
This formula contains the manifold constraint from $\mathcal B$, that is
\[
\tanh\theta=\frac{\beta}{\alpha}=-\frac{V^c r_c}{N}
\]
Likewise
\begin{equation}
\sigma_a^c V^b dr_{cb}=-\beta \bar N\sigma_a^c\nabla_c\ln M
\end{equation}
Therefore we have
\begin{equation} \begin{split}
\sigma_a^c t^b dr_{cb}=&\beta \bar N\sigma_a^c\nabla_c\ln\frac{\bar M}{MN}-N\sigma_a^c\nabla_c\theta\\
=&-\beta\sigma_a^c\nabla_c \bar N-N\sigma_a^c\nabla_c \theta\\
=&-\tilde D_a(\beta \bar N)
\end{split} \end{equation}

(\ref{d^d=d^}) seems to contain some implicitness to express the arbitrariness for foliation selection in action formalism because the 2-form structured by $\sigma_a^c t^b dr_{cb}$ always contributes nothing to integration over $\partial\Sigma_t$ whereas by $\sigma_a^c t^b du_{cb}$ the quantity itself cancels by integrability of $\mathcal B$.

\section{The Hamiltonian constraint} \label{4.THC}
\subsection{Dirac conjecture} \label{sub.Dc}
The Dirac conjecture (see \cite{Dirac1964}) say that any first class secondary constraints should be dynamics independent, in other words, any canonical transformations on it would lead no alteration in physical status, or Hamiltonian vector field of this transformation is independent of Hamiltonian vector field of time-flow.

From the result of \ref{s:ntc} we are able to read the Hamiltonian by varying action under non-equal time variation, thus for gravitational action, its variation write as (\ref{vi}) or (\ref{vit}), which is also viewed as generating function in \cite{Arnowitt2004} if cancel the integral with respect to $t$. Now the Hamiltonian shall be the terms proportional to $\delta t$ at time ``boundary'', that is to say
\begin{equation} \label{HC}
H=\int_{\Sigma_t}N\mathcal H d^3x+2\int_{\partial\Sigma_t}\left(\frac{r_aV_bP^{ab}}{\sqrt{h}}-Nk\right)\sqrt{\sigma}d^2x
\end{equation}
and the generating function a few divergence within \cite{Arnowitt2004},
\begin{equation} \label{GF}
\mathcal G=\int_{\Sigma_t}\left(P^{ab}\delta h_{ab}-N\mathcal H\right)d^3x
\end{equation}
Note that we have accounted of $\delta \{x(t)\}$, the variation for spatial coordinates, which have been absorb in Lie derivative with tensor density via \ref{sub.Td}. On the other hand, the counterpart of generationg function which decide equations in (\ref{vi}) or (\ref{vit}) shall correspond to the same terms and variational coefficients.\footnote{One may refer to \cite{Brown2002} about diffeomorphism invariance of the Hilbert action then will find that momentum constraint always vanishes by Bianchi identity if the vector field leading inner variation tangent to endpoint $\Sigma_t$. Remark that the term with respect to momentum constraint in variation integrates to $\partial\Sigma_t$ and Hamiltonian constraint to endpoint $\Sigma_t$, but only the latter possesses the evolution equation in variation formalism!}

Back to (\ref{HC}) and (\ref{GF}), obviously, there is Hamiltonian constraint but no momentum constraint in both as if the momentum constraint had been removed. Furthermore, a equation or quantity could be viewed as independent by time if all the equations it satisfies are time independent, otherwise, it should be viewed as time dependent. Now as far as we can see from (\ref{vi}) or (\ref{vit}), the Hamiltonian constraint changes by the gravitational system motion because its evolution equation exists in variation formalism under non-equal time variation. We should take notice of the equation as well that $\frac{dH}{dt}+\frac{\partial L}{\partial t}=0$ contains Lagrangian equation $\dot p-\frac{\partial L}{\partial q}=0$ by using the $H=p\dot q-L$, since it represents the fact that all evolutions term in variation formalism should be treated as equations expressing evolutions about configurations as the Lagrangian equations, namely, although the value in $\mathcal H$ no change by transformation, the $\pounds_{\vec t}\mathcal H$ would always contain the system physical status information. Consider transformation $\epsilon\mathcal H$, here $\epsilon=\delta N$, for the system, then remind the fact that the Hamiltonian constraint contains in two equations which in evolution term become $\mathcal H+\epsilon\mathcal H$ by the transformation, further, hold the symmetry in formalism in variation so that $\pounds_{\vec t+\epsilon\vec n}\left(\mathcal H+\epsilon\mathcal H\right)$, which means a transformation towards time that $\vec t \to \vec t+\epsilon \vec n$.

In order to make those transformations generated from $\frac{\partial}{\partial p}$ appear since the Legendre transformations they correspond to are always absent in variation formalism, we can use canonical action principle which cancels all Legendre transformations, the relations between momenta and velocities. Then we just add the $\frac{\partial H}{\partial p}$ terms to (\ref{eq:ntc}) and replace the Lagrangian with Hamiltonian to achieve this goal
\begin{equation} \label{VCA}
\delta I_P=p\delta q-H\delta t+\int\left[\left(\dot q-\frac{\partial H}{\partial p}\right)\delta p-\left(\dot p+\frac{\partial H}{\partial q}\right)\delta q+\left(\frac{dH}{dt}-\frac{\partial H}{\partial t}\right)\delta t\right]dt
\end{equation}
Where $\displaystyle I_P=\int\left(pdq-Hdt\right)$ denotes the canonical action defined in phase space to distinguish $I$ in configurational space. To remind there the phase space indicates augmented phase space $(q, p, t)$ where exists at least one vector fields $\vec v$ to make action 1-form satisfy $v^bd\mathscr L_{ab}=0$ which in Poincar\'e-Cartan formalism $\mathscr L_a=pdq_a-Hdt_a$, or related to symplectic form by $d\mathscr L_{ab}=\Omega_{ab}$ (see \cite{Arnold1989}). Now the variation of canonical action with respect to (\ref{a}) is
\begin{equation} \begin{split} \label{VCGA}
\delta I=&\int dt\int_{\Sigma_t}\bigg[\left(\dot{h}_{ab}-\frac{\delta H}{\delta P^{ab}}\right)\delta P^{ab}+\delta t\pounds_{\vec t}\left(N\mathcal H\right)\\
&-\left(\dot{P}^{ab}+\frac{\delta H}{\delta h_{ab}}\right)\overline{\delta h_{ab}}-\frac{\delta H}{\delta V_a}Nn^b\delta h_{ab}-\frac{\delta H}{\delta N}\delta N\bigg]d^3x\\
&+\int_{\Sigma_{t_0}}^{\Sigma_{t_1}}\left(P^{ab}\delta h_{ab}-N\mathcal H\delta t\right)d^3x+\{\textrm{boundary term}\}\\
\end{split} \end{equation}
Because of absence in physical sense as we have known, (\ref{VCGA}) removes the primary constraints for the sake of abridging. Now consider Hamiltonian $H_E=H+\mathcal C_{\vec v}$ where uses the term in \cite{Dirac1964} that denotes $H_E$ as extended Hamiltonian, and $H$, also in others place, as an abbreviation for total Hamiltonian $H_T$, and $\mathcal C_{\vec v}=\displaystyle\int_{\Sigma_t}v^a\mathcal H_ad^3x$. Demand that the $v^a$ should vanish on boundary $\Sigma_t$ as taking it into consideration demanding no contribution in value in Hamiltonian to which constraint functional leads. To hold the canonical equations in (\ref{VCGA}), compute the (the Hamilton theory about this see \cite{Arnold1989,Arnowitt2004,Ashtekar1988})
\begin{align*}
\frac{\delta H_E}{\delta P^{ab}}&=\frac{\delta H}{\delta P^{ab}}+\frac{\delta\mathcal C_{\vec v}}{\delta P^{ab}}=\pounds_{\vec t+\vec v}h_{ab}\\
-\frac{\delta H_E}{\delta h_{ab}}&=-\frac{\delta H}{\delta h_{ab}}+\frac{\delta\mathcal C_{\vec v}}{\delta h_{ab}}=\pounds_{\vec t+\vec v}P^{ab}\\
\end{align*}
However, the
\begin{align*}
\frac{\delta H_E}{\delta V_a}&=\frac{\delta H}{\delta V_a}\\
\frac{\delta H_E}{\delta N}&=\frac{\delta H}{\delta N}
\end{align*}
In terms of symplectic manifold on phase space, the Hamiltonian vector field of $\mathcal C_{\vec v}$ leads to a transformation $h_{ab} \to \phi_{\vec v}^*h_{ab}$ and $P^{ab} \to \phi_{\vec v}^*P^{ab}$ like a infinitesimal shift vector.

Then analyze the Hamiltonian constraint $\mathcal H$. Noting there are two equations relevant to this constraint which we may view them as Lagrangian equation with respect to $\dot{P}_N$ and $\ddot{P}_N$ a quadratic differential equation with time. To extend the Hamiltonian with Hamiltonian constraint, the generating function is
\[
\mathcal C_{\epsilon}=\int_{\Sigma_t}\epsilon \mathcal Hd^3x
\]
It acts as a Hamiltonian without shift vector. According the argument had done in \ref{DI} that $\frac{dH}{dt}-\frac{\partial H}{\partial t}=0$ contains canonical equations, the equation $\pounds_{\vec t}\left(N\mathcal H\right)$ in (\ref{VCGA}) represents a canonical equation itself as well. To ignore boundary term, it writes as
\[
\frac{\delta \left(N\mathcal H\right)}{\delta h_{ab}}\pounds_{\vec t}\,h_{ab}+\frac{\delta \left(N\mathcal H\right)}{\delta P^{ab}}\pounds_{\vec t}P^{ab}=0
\]
As the formula above, equation $\pounds_{\vec t}\left(N\mathcal H\right)$ describes the gravitational system evolution flowing by time with lapse function $N$ that $\frac{\delta \left(N\mathcal H\right)}{\delta h_{ab}}$ and $\frac{\delta \left(N\mathcal H\right)}{\delta P^{ab}}$ will come out. Suppose we put a Hamiltonian constraint proportional to $\epsilon$ into Hamiltonian constraint, that we will get a extended Hamiltonian $\left(N+\epsilon\right)\mathcal H$, the variables in phase space would alter with $\frac{\delta \left(\epsilon\mathcal H\right)}{\delta h_{ab}}$ and $\frac{\delta \left(\epsilon\mathcal H\right)}{\delta P^{ab}}$, equivalent to lapse function altered with $\epsilon$. In fact, simpler from $\pounds_{\vec t}\left(N\mathcal H\right)$ viewpoint directly, considering the formalism invariant in variation expression we have $\pounds_{\vec t}\left(N\mathcal H\right) \to \pounds_{\vec t+\epsilon \vec n}\left(N\mathcal H+\epsilon\mathcal H\right)$ after transformation done.

To recall how Hamiltonian constraint came out from variation we may perceive the analogy to arising of Hamiltonian in (\ref{eq:ntc}) that
\[
2N\left(K^2-K^{ab}K_{ab}\right)\pounds_{\vec t}\,\delta t
\]
derived from variation $\delta\left(K^{ab}K_{ab}-K^2\right)$ is analogous to the term $-p\dot{q}\frac{d(\delta t)}{dt}$ and the ``$p\dot q$'' is ``true dynamical degrees of freedom'' which expresses as $2N\left(K^{ab}K_{ab}-K^2\right)$ in the case of gravitation as we have already known. The term $Nn^c\delta n_c$ leads to the Hamiltonian constraint and its evolution simultaneously since it split as $-\delta N-N\pounds_{\vec t}\,\delta t$ therefore it is hard to suppose the situation where the $\delta t\pounds_{\vec t}\left(N\mathcal H\right)$ is removed but the $\mathcal H\delta N$ remained. The Significance exists for hardness may arise again on the discussion of \ref{sub.R-p}. Remark ``$p\dot q$'' in total Hamiltonian should be
\[
P^{ab}\dot h_{ab}=2N\left(K^{ab}K_{ab}-K^2\right)+2P^{ab}D_a V_b
\]
I think it implies reduction of the momentum constraint in Hamiltonian on the other hand since the last term with respect to momentum constraint doesn't appear in ``$p\dot q$'' in variation formalism.

The Poincar\'e-Cartan integral invariant formalism may offer a more distinct explanation to reveal difference between the situations that the $\pounds_{\vec t}\left(N\mathcal H\right)$ removed or existed. Applying the Hamilton-Jacobi analysis, the action whose system obeys geometrodynamics equations satisfies
\begin{equation} \label{GA P-C}
\delta I=\int_{\Sigma_t}\left(P^{ab}\delta h_{ab}-N\mathcal H\delta t\right)d^3x
\end{equation}
As well as we have known former, the momentum constraint removed but Hamiltonian constraint existed. Analogous to analysis in \ref{s:ntc}, we have Poincar\'e-Cartan invariance
\begin{equation}
\oint_{C}\int_{\Sigma_t}\left(P^{ab}\delta h_{ab}-N\mathcal H\delta t\right)d^3x
\end{equation}
Where the concept of phase space expanded that the $h_{ab}$ and $P^{ab}$ treated as canonical coordinate and each functional action represented a point with respect to its corresponding system in the phase space. If the $\pounds_{\vec t}\left(N\mathcal H\right)$ was removed from variation, its integration over time would vanish simultaneously, then the Poincar\'e-Cartan invariance would be $\displaystyle \oint_{C}\int_{\Sigma_t}P^{ab}\delta h_{ab}d^3x$. In this case, the integral is more unconstrained than the integral added with $\displaystyle\int_{\Sigma_t}N\mathcal H\delta t$, since the former without introduction by ``abstract'' vector $\left(\delta h_{ab}, \delta t\right)$ hence the divergence in $\delta t$ canceled among two point $\delta h_{ab}$ given. That is to say that the Hamiltonian constraint transforms freely which might represents a parameter translation merely, yet the state fixed if the integral invariance is (\ref{GA P-C}).

In brief, the change to integral circuit holds functional ``$0$'' if the Hamiltonian constraint makes no difference to physics state whereas only holds numerical ``$0$'' if makes difference to physics state.

The key of these analyses towards reduced Hamiltonian focuses on the term proportional to $\delta$ in variational action. The gravitational action owns the term $N\mathcal H$ and it associated with the configuration related to time. We are able to write the reduced Hamiltonian in Maxwell system that Lagrangian $\mathscr L=F_{\mu\nu}F^{\mu\nu}$ on Minkowski spacetime. Now if demand $\delta t \neq 0, \delta x^i=0$ and $\partial_i \delta t=0$ we will have
\[
\delta I_{\scriptscriptstyle EM}=\int\left[4F^{\mu\nu}\partial_{\mu}\delta A_{\nu}-\left(2F_{0i}F^{0i}-F_{ij}F^{ij}\right)\partial_t \delta t\right]dtd^3x
\]
Easy to read the reduced Hamiltonian density
\[
\mathscr H_{\text re}=2F_{0i}F^{0i}-F_{ij}F^{ij}
\]
compared with $\mathscr H=\mathscr H_{\text re}+4F_{0i}\partial^i \varphi$ by $H=p\dot q-L$ which owns the constraint $\partial^i F_{0i}$ removed in $\mathscr H_{\text re}$.

\subsection{Re-parametrization} \label{sub.R-p}
As we have see before, the $\mathcal H$ ought not be considered as a removable constraint like momentum constraint $\mathcal H_a$, while represents a ``true dynamical degrees of freedom'' (see \cite{Wald1984}). From the \cite{Arnowitt2004} we can see the general relativity as ``already'' parameterized theory. There is possible to cancel Hamiltonian constraint with re-parametrization treatment in some case such as particle action where we call $\tau$ as the ``new'' time parameter used to substitute the $t$ satisfied $dt=\frac{dt}{d\tau}d\tau$ thus $t$ to be a configurational coordinate conjugated its general momentum $-H$. In \ref{s:ntc} the variation of re-parameter action has been discussed with configurational action principle and now we discuss the canonical action principle. There are momenta conjugated its coordinates (see \cite{Arnowitt2004})
\begin{equation} \begin{split} \label{pRP}
p_q&=\frac{\partial \left(\dot{t}L\right)}{\partial \dot{q}}=p\\
p_t&=\frac{\partial \left(\dot{t}L\right)}{\partial \dot{t}}=L-p\frac{dq}{dt}=-H_t
\end{split} \end{equation}
Where $\dot t=\frac{dt}{d\tau}$ and $H_t$ is Hamiltonian for the system with parameter $t$ then the system with parameter $\tau$ has a primary constraint
\[
\mathcal C_t=p_t+H_t\left(q,p,t\right)=0
\]

There is a map from $\Gamma_t$ to $\Gamma$ which views $t$ as a coordinate, that
\begin{align*}
L\colon\Gamma_t &\to \Gamma\\
\mathcal C_t\colon\Gamma_t& \to 0
\end{align*}
Denote $\Gamma_{\tau}\equiv L(\Gamma_t)$ and naturally give a map $\mathcal C_t\colon\Gamma_{\tau} \to 0$. Push forward of map gives $L_*\vec t=\vec \tau \subset V_{\Gamma_{\tau}}$ since we have $t=t(\tau)$. About Hamiltonian we define $H_{\Gamma}\equiv f\mathcal C_t=f\left(p_t+H_t\right)$. No necessary to demand each point on $\Gamma$ satisfied $H_{\Gamma}=0$ or $\mathcal C_t=0$, but let it be zero on $\Gamma_{\tau}$. Now the canonical action with parameter $\tau$ written as
\begin{equation} \label{CPA}
I_P(\tau)=\int \left(p\dot q+p_t\dot t-H_{\Gamma}\right)d\tau
\end{equation}
Its variation is
\begin{equation} \begin{split} \label{VCPA1}
\delta I_P(\tau)=&p\delta q+p_t\delta t+\int \bigg[\left(\dot{q}-\frac{\partial H_{\Gamma}}{\partial p}\right)\delta p-\left(\dot{p}+\frac{\partial H_{\Gamma}}{\partial q}\right)\delta q\\
&+\left(\dot{t}-\frac{\partial H_{\Gamma}}{\partial p_t}\right)\delta p_t-\left(\dot{p}_t+\frac{\partial H_{\Gamma}}{\partial t}\right)\delta t\bigg]d\tau-\int H_{\Gamma}d(\delta\tau)
\end{split} \end{equation}
The total Hamiltonian $H_{\tau}=f\mathcal C_t$, to put it into (\ref{VCPA1}) and split it, we obtain
\begin{equation} \begin{split} \label{VCPA2}
\delta I_P(\tau)=&p\delta q+p_t\delta t+\int \bigg[\left(\dot{q}-f\frac{\partial H_t}{\partial p}\right)\delta p-\left(\dot{p}+f\frac{\partial H_t}{\partial q}\right)\delta q\\
&+\left(\dot{t}-f\frac{\partial H_t}{\partial p_t}\right)\delta p_t-\left(\dot{p}_t+f\frac{\partial H_t}{\partial t}\right)\delta t\bigg]d\tau\\
&-\int \left[\mathcal C_t\delta fd\tau+f\mathcal C_td(\delta\tau)\right]
\end{split} \end{equation}
Very analogous to (\ref{VCGA}) where the $f$ corresponds with the $N$ and $\mathcal C_t$ with $\mathcal H$ respectively, yet the divergence in primary constraint with secondary constraint.

According to the expression of $p_t$, the $H_{\Gamma}$ ought be canceled hence $I_P(\tau)=I_P(t)$. However, there is no bridge between $\dot q$ and $p$ in phase space due to absence of Legendre transformation when apply the canonical action principle and that is why we could not declare the $I=I_P$ without Legendre transformation (ie: More direct discussion reveals on path integral approach about two action). Yet if we declare the two system equivalent described by $t$ or $\tau$, the Legendre transformation will be satisfied due to equality in their Lagrangian. Actually, there is not evidence to say that both of the two parameters describe the same system if neither Hamiltonian nor Lagrangian is given, or in other word, the expression of $H_{\Gamma}$ will tell us if the $(t, p_t)$ can be chosen as a dynamical parameter. As far as these are concerned, the equation $p_t=\frac{\partial \left(\dot{t}L\right)}{\partial \dot{t}}=L-p\frac{dq}{dt}$ is applied hence the $H_{\Gamma}=0$ with its definition thus the canonical action (\ref{CPA}) with $\tau$
\[
I_P(\tau)=\int \left(p dq-H_t dt\right)=I_P(t)
\]
Obviously, $\delta I_P(\tau)$ must be identical to $\delta I_P(t)$ as $I(\tau)$ to $I(t)$. The condition that both of the two parameters describe the same system also represents as $L_*\vec t=\vec \tau \subset V_{\Gamma_{\tau}}$. To recall the last term in (\ref{VCPA2}), define the substitution $dT=fd\tau$ then
\[
-\int\mathcal C_t d\left(\delta T\right)=-\mathcal C_t\delta T+\int \delta T\pounds_{\vec T}\mathcal C_t dT
\]
this term canceled since $L(\Gamma_t) \in \Gamma_{\tau}$ and $\vec \tau$ tangent $\Gamma_{\tau}$.

From arguments above, we are able to cancel the term $\mathcal C_t \delta\left(fd\tau\right)$ for entire $\Gamma_t$, so is $\tau$, no matter whether canonical equations satisfied or deviated. It is the cancellation that demonstrates the equivalence in variation between canonical action form and configurational action, moreover, the Hamiltonian principle in configurational space apart from Legendre transformation terms. More detail about it
\begin{equation*} \begin{split}
\delta I=&p\delta q-H\delta t+\int\left[\left(\frac{\partial L}{\partial q}-\frac{dp}{dt}\right)\delta q+\left(\frac{\partial L}{\partial t}-\frac{dp_t}{dt}\right)\delta t\right]dt\\
&+\int\left(L\dot t-p_t\dot t-p\dot q\right)d(\delta \tau)
\end{split} \end{equation*}
The last term always be canceled by the definition of Hamiltonian and in this formalism. Therefore we are able to conclude that this formalism is de-parameterized itself. Therefore, let us concentrate attention on configurational action variation since $\tau$ always absent clearly in this action principle forms.

At first we denote
\[
n_c(T)=-\nabla_c T
\]
and
\[
\mathcal H(T)=-2G_{ab}n^a(T)n^b(T)\sqrt{h(T)}
\]
This is the Hamiltonian constraint of $\Sigma_T$ satisfied $\sqrt{-g(T)}=\sqrt{h(T)}$ for $N(T)=1$ where $\sqrt{-g(T)}$ denotes the determinant of $g_{ab}$ under the coordinates basis $\{x_{\scriptscriptstyle T},T\}$. Two parameter $t,T$ are related by the $\nabla_cT=N\nabla_ct$ and there I have made a hypothesis that we are able to splice the spacetime $\mathcal M$ to a hypersurfaces family $\Sigma_T$ where each point satisfied $g^{ab}\nabla_a T\nabla_b T=-1$. About this hypothesis, there are some discussions following:\\*
1.In most of case, there is no possibility to achieve this hypothesis if the $\Sigma_{t_0}$ and $\Sigma_{t_1}$ were fixed;\\*
2.Start with $\Sigma_{t_0}$ to slice $\mathcal M$ by $g^{ab}\nabla_a T\nabla_b T=-1$, and then leads to some leaves with corner when slice $\Sigma_{t_1}$. However, it may turn space-like vector fields to time-flow vector.

For overcoming these difficulties, recall the (\ref{VG}) derived from (\ref{vi}), then rewrite the (\ref{VG})
\begin{equation}
\int_{\mathcal M}G_{ab}\delta g^{ab}\sqrt{-g(T)}dTd^3x_{\scriptscriptstyle T}
\end{equation}
the Hamiltonian constraint in variation thus writes as
\begin{equation}
\int dT\int_{\Sigma_T}\mathcal H(T) n^c(T) \delta n_c(T) d^3x_{\scriptscriptstyle T}
\end{equation}
No matter how we slice the $\mathcal M$ the re-parametrization ought be achieved locally at least. Or, on a special side, we are only meant to discuss those $\mathcal M=\cup_T\Sigma_{t_0}$. Anyhow, it is possible to write the Hamiltonian constraint with $\{x_{\scriptscriptstyle T},T\}$ in variation that
\begin{equation} \label{RP}
-\int\!\!\!\int_{\Sigma_T}\mathcal H(T) d\left(\delta T\right)d^3x_{\scriptscriptstyle T}=-\int_{\Sigma_{T_0}}^{\Sigma_{T_1}}\mathcal H\delta T+\int dT\int_{\Sigma_T}\pounds_{\vec T}\mathcal H(T)d^3x_{\scriptscriptstyle T}
\end{equation}
Identical consequence can be derived from (\ref{vi}) of course, which also reflects the formalism invariance in variation. Perhaps this consequence can be viewed as $N(T)=1$\footnote{It is one of ``Imposition of coordinate conditions'' in \cite{Arnowitt2004}.} fixed as well when vary it we have $\delta N(T)=0$.

As we have seen in (\ref{RP}), even though the $\mathcal H$ in variation removed by re-parametrization, we are not able to remove its evolution expressed as another equation
\[
\pounds_{\vec T}\mathcal H(T)=0
\]
From the view of configurational action principle the variation have not de-parameterized for the existence of $\pounds_{\vec T}\mathcal H(T)$ which is not a Legendre transformation. Moreover, at the view of canonical action principle, for the the mechanics system of (\ref{pRP}), the map is from $\Gamma_t$ to $\Gamma_{\tau}$ that cancels the $\mathcal C_t$ in variation because where each point satisfies Legendre transformation automatically including those points deviated canonical equations. Whereas it must be still on secondary constraint surface for each point that we were able to cancel the $\pounds_{\vec T}\mathcal H(T)=0$ then insist the $\mathcal H$ is removable in case of gravitation under mapping from $\Gamma_t$ to $\Gamma_T$, not as the case of (\ref{VCPA2}) where each point are still on primary constraint surface automatically. Yet it would be equivalent to stick the existence of Hamiltonian constraint $\mathcal H$. Following all the considerations of these arguments we shall face the fact that gravitational system are not able to be de-parameterized (see \cite{Ashtekar1988}). It seems to impose us that the $\mathcal H$ is not only constraint but also the true Hamiltonian.

\end{document}